\documentclass[a4paper,11pt]{article}
\usepackage{aaskaiid}
\usepackage{aas_macros}
\usepackage{comment}
\usepackage{orcidlink}
\title{Supernova remnants in the new radio astronomy era}
\ShortTitle{Supernova remnants}

\author[1]{A.~Ingallinera\orcidlink{0000-0002-3137-473X}}
\ShortName{Adriano Ingallinera et al.} 
\author[2]{M.~Arias\orcidlink{orcid:0000-0002-7918-904X}}
\author[3]{G.~Castelletti\orcidlink{0009-0002-0134-2064}}
\author[1]{C.~Bordiu\orcidlink{0000-0002-7703-0692}}
\author[1]{F.~Bufano\orcidlink{0000-0002-3429-2481}}
\author[4]{G.~Cosentino\orcidlink{0000-0001-5551-9502}}
\author[5]{M.~D.~Filipovi\'c\orcidlink{0000-0002-4990-9288}}
\author[6,7]{D.~Liu\orcidlink{0009-0001-9837-9455}}
\author[1]{S.~Loru\orcidlink{0000-0001-5126-1719}}
\author[8]{S.~Orlando\orcidlink{0000-0003-2836-540X}}
\author[8,9]{O.~Petruk\orcidlink{0000-0003-3487-0349}}
\author[5]{Z.~Smeaton\orcidlink{0009-0009-7061-0553}}
\author[10]{A.~Traficante\orcidlink{0000-0003-1665-6402}}
\author[1]{C.~Trigilio\orcidlink{0000-0002-1216-7831}}
\author[1]{G.~Umana\orcidlink{0000-0002-6972-8388}}
\author[11]{G.~Anderson\orcidlink{0000-0001-6544-8007}}
\author[8]{F.~Bocchino\orcidlink{0000-0002-2321-5616}}
\author[1]{C.~Buemi\orcidlink{0000-0002-7288-4613}}
\author[1]{F.~Cavallaro\orcidlink{0000-0003-1856-6806}}
\author[12]{E.~Egron\orcidlink{0000-0002-1532-4142}}
\author[11]{N.~Hurley-Walker\orcidlink{0000-0002-5119-4808}}
\author[13]{R.~Kothes\orcidlink{0000-0001-5953-0100}}
\author[1]{P.~Leto\orcidlink{0000-0003-4864-2806}}
\author[11]{S.~Mantovanini\orcidlink{0000-0003-1210-5603}}
\author[14,8]{M.~Miceli\orcidlink{0000-0003-0876-8391}}
\author[15]{G.~Morlino\orcidlink{0000-0002-5014-4817}}
\author[12]{A.~Pellizoni\orcidlink{0000-0002-4590-0040}}
\author[16]{M.~Sasaki\orcidlink{0000-0001-5302-1866}}

\affiliation[1]{INAF - Osservatorio Astrofisico di Catania, 95123 Catania, Italy}
\emailAdd{adriano.ingallinera@inaf.it}
\affiliation[2]{Instituto de Astrofisica de Andalucia (IAA-CSIC), Glorieta de la Astronomia s/n, 18008 Granada, Spain}
\affiliation[3]{Instituto de Astronomía y Física del Espacio (IAFE, UBA-CONICET), Av. Int. Güiraldes 2620, Pabellón IAFE, Ciudad Universitaria, 1428 Buenos Aires, Argentina}
\affiliation[4]{Institut de Radioastronomie Millimétrique, 300 Rue de la Piscine, 38400, Saint-Martin-d'Hères, France}
\affiliation[5]{Western Sydney University, Locked Bag 1797, Penrith South DC, NSW 2751, Australia}
\affiliation[6]{College of Science, China Three Gorges University, Yichang 443002, China}
\affiliation[7]{Center for Astronomy and Space Sciences, China Three Gorges University, Yichang 443002, China}
\affiliation[8]{INAF - Osservatorio Astronomico di Palermo, Piazza del Parlamento 1, 90134 Palermo, Italy}
\affiliation[9]{Institute for Applied Problems in Mechanics and Mathematics, National Academy of Sciences of Ukraine, Naukova St. 3-b, 79060 Lviv, Ukraine} 
\affiliation[10]{INAF, IAPS, Via Fosso del Cavaliere, 100, 00133 Rome, Italy}
\affiliation[11]{International Centre for Radio Astronomy Research, Curtin University, Bentley WA 6102, Australia}
\affiliation[12]{INAF - Osservatorio Astronomico di Cagliari, Via della Scienza 5, 09047 Selargius, Italy}
\affiliation[13]{Dominion Radio Astrophysical Observatory, Herzberg Astronomy and Astrophysics, National Research Council Canada, PO Box 248, Penticton, BC V2A 6J9, Canada}
\affiliation[14]{Dipartimento di Fisica e Chimica E. Segrè, Università degli Studi di Palermo, Via Archirafi 36, 90123 Palermo, Italy}
\affiliation[15]{INAF, Osservatorio Astrofisico di Arcetri, L.go E. Fermi 5, Firenze, 50125, Italy}
\affiliation[16]{Dr. Karl Remeis Observatory, Erlangen Centre for Astroparticle Physics, Friedrich-Alexander-Universität Erlangen-Nürnberg, Sternwartstr. 7, 96049, Bamberg, Germany}

\newcommand{\red}[1]{\textcolor{red}{#1}}

\abstract{
Supernova remnants (SNRs) are what is left after stellar explosions, when the stellar ejecta, the explosion shock and the circumstellar medium interact. Despite being among the first objects studied in radio astronomy, observational difficulties have so far prevented a definitive characterisation, which would help answer open questions related to these sources. It is debated which is the contribution of SNRs to Galactic cosmic rays, or how the interaction with the surrounding environments influences the particle energetics. The SKA precursors are providing valuable and unexpected discoveries on SNRs, thanks to their unique capabilities to probe spatial scales from a few arcseconds to a few degrees with a sensitivity of tens of microjansky. Accurate integrated flux density measurements and arcsecond-scale spectral-index maps are now possible for tens of SNRs, substantially expanding the small subset of remnants traditionally studied in great detail. SKA will markedly enhance current observations by providing: higher sensitivity, enabling the detection of fainter SNRs also in polarisation, revealing diffuse structures and the underlying magnetic field configuration; higher angular resolution, allowing detailed mapping of compact remnants and reducing depolarisation in fine structures, tracing filaments and shocks fronts; wider frequency coverage to probe unexplored spectral windows, where spectral turnovers and breaks or cut-off may occur, establishing a direct connection to X-ray and $\gamma$-ray emission that constrains the electron population, and enabling accurate modelling of the non-thermal emission across the electromagnetic spectrum; improved image fidelity for more reliable cross-matching with other wavelengths, leading to a better understanding of the SNR-interstellar medium interplay.
}

\begin{document}
\maketitle

\include{journal-names}

\section{Introduction}

Supernova remnants (SNRs) are produced after stellar explosions, when complex interactions arise between the stellar ejecta, the explosion shock, the magnetic field, and both the circumstellar (CSM) and interstellar medium (ISM). Because of their non-thermal emission in the radio band, they are prominent objects in the Galactic plane, and they have been studied since the very beginning of radio astronomy. 
However, radio observations of SNRs have long faced a range of technical challenges, often requiring compromises that have, for most cases, prevented an accurate characterisation of their emission properties (brightness, spectral behaviour, polarisation) and morphology (low-brightness diffuse features missed by interferometers, or very small-scale structures unresolved by single-dish telescopes). Such a characterisation is critical for a comprehensive understanding of the nature, evolution, and role of SNRs in the life cycle of matter in our Galaxy. Indeed, many open questions remain, among them:
\begin{itemize}
    \item 
    it remains uncertain whether the current number of known or candidate Galactic SNRs--at least a factor of 2--3 lower than expected from stellar evolution models--reflects observational biases or an incomplete census \citep{Brogan_2006,2015A&ARv..23....3D};
    \item 
    it is still debated what fraction of Galactic cosmic rays (CRs) up to $\sim\!1$~PeV originates in SNRs, as other sources such as pulsar wind nebulae 
    and superbubbles may contribute comparably \citep{Cristofari2021};
    \item it is not clear how interactions with the surrounding environments influence particle energetics \citep{Cristofari_2021A&A};
    \item it is poorly understood how SNRs affect star formation in terms of chemical and mechanical feedback.
\end{itemize}

As soon as the SKA precursors became operational, new observations, especially from large-area surveys, have significantly contributed to our knowledge of SNRs, in our Galaxy, in the Magellanic Clouds, and also in other galaxies. In this context, ASKAP, MeerKAT, and MWA are all proving extremely valuable,
providing accurate measurements for a large sample of SNRs and enabling the discovery of new ones. Their combined strength lies in the fact that, by bringing together the complementary capabilities of all three instruments, they offer unique advantages such as: 
\begin{itemize}
    \item probing spatial scale from a few arcseconds to a few degrees;
    \item reaching a point-source sensitivity down to $\sim\!10$~µJy~beam$^{-1}$;
    \item offering a continuous frequency coverage from $\sim\!80$~MHz to $\sim\!3.5$~GHz.

\end{itemize}

\begin{figure}
    \centering
    \includegraphics[width=0.85\linewidth]{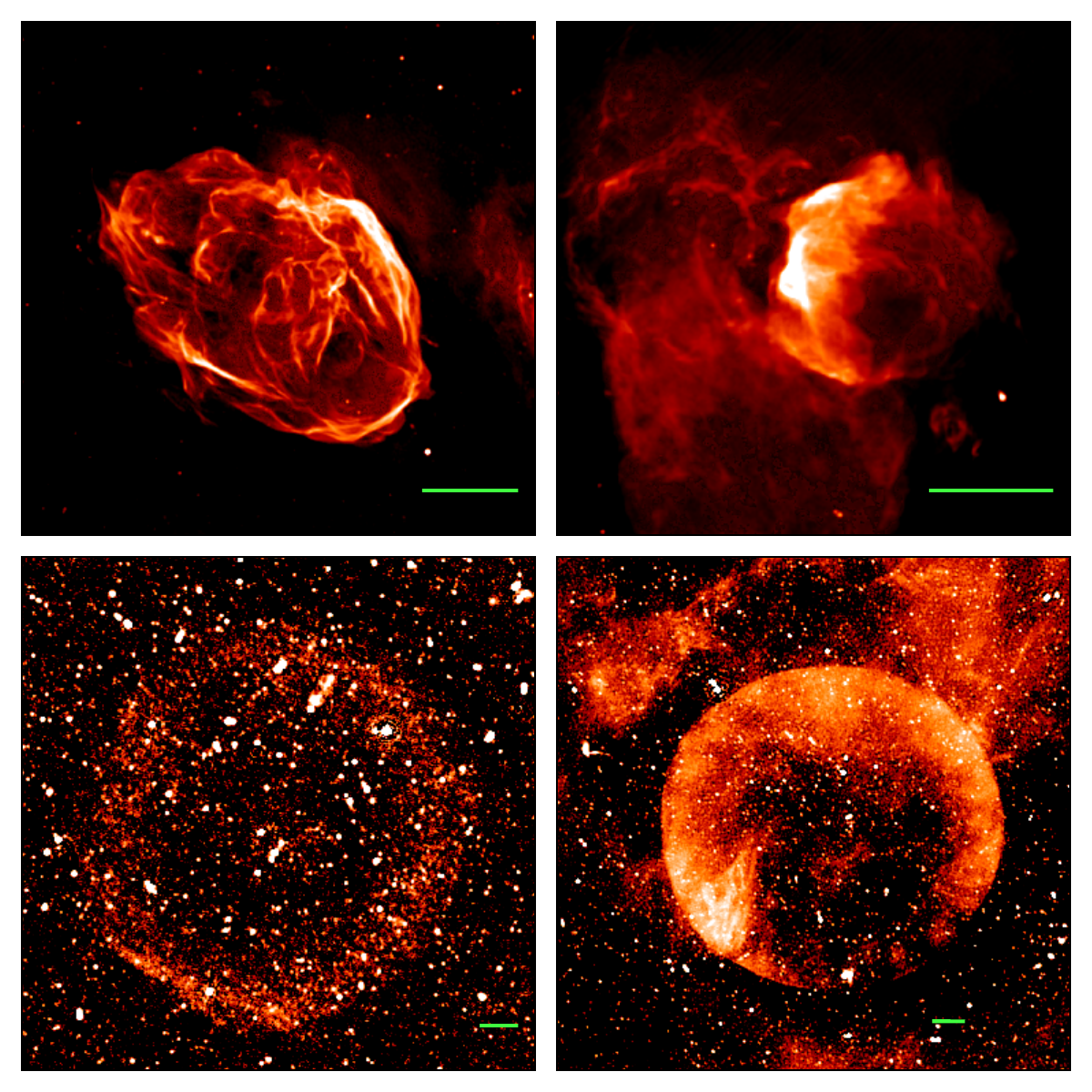}
    \caption{a gallery of SNRs and SNR candidates recently observed with SKA precursors and pathfinders. (\textit{top}) MeerKAT images of SNR~G296.8-00.3 and SNR~G348.7+00.3 at 1.3~GHz \citep{Loru2024}. (\textit{bottom left}) The low surface brightness Calvera's SNR from the LOFAR Two-Metre Sky Survey at 144~MHz \citep{Shimwell2017}, discovered by \citet{Arias2022}, and proposed to result from the explosion of a massive runaway star that also formed the Calvera pulsar \citep{Dakic2025,Greco2025}. (\textit{bottom right}) The Teleios SNR candidate discovered in EMU images at 0.9~GHz \citep{Filipovic2025}. The green scale bar is 5-arcmin long in all the four panels.}
    \label{fig:gallery}
\end{figure}

Large radio surveys (e.g.,  EMU\footnote{Evolutionary Map of the Universe (EMU; \citealt{Norris_2021})} and POSSUM\footnote{POlarisation Sky Survey of the Universe's Magnetism (POSSUM; \citealt{Gaensler_2010})} with ASKAP, SMGPS\footnote{SARAO MeerKAT legacy Galactic
Plane Survey (SMGPS; \citealt{Goedhart_2024})} with MeerKAT, GLEAM\footnote{GaLactic and Extragalactic All-sky MWA (GLEAM; \citealt{Wayth_2015})} with MWA) are offering remarkable 
opportunities in this field, as evidenced by the 
numerous papers published 
to date. New SNR candidates have been identified, more than doubling the number of possible Galactic SNRs (e.g.,  \citealt{Hurley-Walker2019}; \citealt{Ball2023}; \citealt{Lazarevic2024}; \citealt{Smeaton2024}; \citealt{Anderson2025}), and have also significantly improved our knowledge of SNRs in the Magellanic Clouds (e.g.,  \citealt{Bozzetto2023}; \citealt{Cotton2024}).  Figure \ref{fig:gallery} shows an example of known and new SNRs recently imaged with SKA precursors and pathfinders.
Furthermore, the characterisation of spectral energy distributions (SEDs), the production of spectral-index maps, and the analysis of morphological features have been critically examined or presented for the first time (e.g.,  \citealt{Castelletti2021}; \citealt{Bufano2024}; \citealt{Loru2024}; \citealt{Filipovic2024}).

One of the main opportunities enabled by the SKA precursors is that accurate integrated flux density measurements and arcsecond-scale spectral index maps are now feasible for tens of SNRs, particularly in the previously poorly explored southern hemisphere. This allows us to substantially enlarge the small subset of SNRs that have traditionally been studied in great detail (such as Cas~A, Tycho, Kepler). Although these well-known remnants have so far dominated our understanding of Galactic SNRs and have been crucial in shaping the current theoretical framework, they may also have introduced a selection bias that SKA observations can finally overcome.

SKA-LOW and MID will significantly improve the current observations.
\begin{itemize}
    \item 
    The higher sensitivity compared to existing instruments will be crucial to: (i) discover new, low-surface-brightness remnants; (ii) characterise fainter SNRs by tracing diffuse emission components; and (iii) perform polarimetric studies, unveiling the magnetic field configuration.
    \item 
    The arcsecond and sub-arcsecond angular resolution will enable (i) detailed mapping of compact SNRs; (ii) reduced depolarisation in finer structures, greatly improving the study of filaments and shock fronts; and (iii) analyses of the power spectra of the turbulent component of the magnetic field.
    \item 
    The broad frequency coverage, from 50~MHz to 15~GHz, will open unexplored spectral windows where turnovers (SKA-LOW, $<330$~MHz) and spectral breaks or cut-offs  (SKA-MID, $>330$~MHz) are expected to occur. This will provide key constraints to the electron population and allow for accurate modelling of the non-thermal emission across the entire electromagnetic spectrum, in connection with X- and $\gamma$-ray data.

    \item 
    The improved $uv$-coverage, resulting from the larger number of antennas, will enhance image fidelity, enabling more reliable multi-wavelength cross-comparisons and a better understanding of the SNR-ISM interplay. 

\end{itemize}

In Section \ref{sec:sci}, we will present some science cases in which observations with SKA will lead to new important discoveries. In Section 3, we will discuss possible commensalities with other SKA cases and synergies with other facilities.

\section{Science cases}
\label{sec:sci}
Since SNRs are intrinsically complex objects, understanding their characteristics and evolution requires a comprehensive approach that combines theoretical modelling with observations. Multiwavelength studies are essential in this context, providing complementary insights into the diverse physical processes that shape these objects. In this Section, we outline several science cases where the SKA will make a transformative contribution beyond what is currently achievable.

\subsection{Radio spectral characterisation}
\label{sec:sed}
Radio emission from SNRs is dominated by synchrotron radiation produced by relativistic electrons. Provided that the electrons 
follow a power-law energy distribution, the resulting radio emission  
also exhibits a power-law spectrum,
with the radio spectral index $\alpha$ directly related to the energy spectral index $s$. Specifically, for an energy distribution of the form
\[
N(E)\propto E^{-s},
\]
the corresponding synchrotron spectrum--which in many cases extends at least up to a few tens of GHz--is given by
\[
S(\nu)\propto\nu^\alpha
\]
with
\[
\alpha=\frac{1-s}{2}.
\]

The diffusive shock acceleration (DSA) theory for test particles predicts an electron energy index $s=2$, which corresponds to a synchrotron spectral index $\alpha=-0.5$ \citep{1983RPPh...46..973D}. 
Observationally, the radio spectral indices 
for many SNRs cluster around this value, although a considerable scatter is observed, with reported values ranging from 0.0 to $-1.1$ \citep{Reynolds_2011, 2015A&ARv..23....3D, Green2025}. Part of this dispersion arises from systematic uncertainties, such as (i) measurements obtained with different instruments; (ii) calibration errors; (iii) the limited frequency coverage (typically from $\approx0.2$ to $\approx5$~GHz); and (iv) differences in the flux extraction regions among studies. 
Nevertheless, even accounting for these effects, it is evident that spectral indices deviating from $-0.5$ are genuine. A clear indication that the spectral index can assume values different from $-0.5$ is the spatial variability of $\alpha$ observed across many remnants, with typical differences of up to $\approx0.3$ between regions \citep{Castelletti2021, Loru2024}. Although the aforementioned sources of uncertainty also affect the derivation of spectral index maps, there is strong evidence that these spatial variations are real, as they often coincide with prominent brightness structures such as filaments (e.g., W44 -- \citealt{Castelletti2007};  IC~443 -- \citealt{Castelletti2011}; Kes~73 -- \citealt{Ingallinera2014}). SNRs can thus be characterised by both a global spectral index, providing synthetic information on the average energy distribution of relativistic electrons, and a local spectral index, whose spatial variations can be traced through spatially resolved spectral index maps \citep{Sun2011, Castelletti2021, Loru2024}.
The spectral index of a SNR is also thought to evolve with time \citep[e.g., in SN~1987A,][]{2010ApJ...710.1515Z}. Since the radio spectral index $\alpha$ directly reflects the slope of the relativistic electron energy distribution, deviations from the canonical value $\alpha=-0.5$ likely arise from specific features of particle acceleration. In particular, the energy index $s$ may be modified during acceleration by several processes: velocity gradients across the shock--both upstream, due to the back-reaction of accelerated particles \citep{2005MNRAS.364L..76A}, and downstream, as a result of the flow hydrodynamics \citep{2024A&A...688A.108P}-- re-acceleration of pre-existing CRs \citep{2004APh....21...45B}, Alfvénic drift \citep{2008AIPC.1085..336Z}, or stochastic re-acceleration of particles immediately downstream \citep{2020A&A...639A.124W}.



Deviations from a pure power-law spectrum are commonly observed at frequencies below 100~MHz, where a turnover can be detected. 
These low-frequency departures are primarily attributed to thermal free-free absorption, which illustrates the intrinsic properties of an SNR, its surrounding medium, or the line-of-sight ISM:
\begin{itemize}
    \item 
    Foreground absorption along the line of sight: The extended and often bright radio emission of Galactic SNRs provides a convenient background against which the ionised gas in the Milky Way can be detected through free-free absorption. Such cases offer valuable probes of the distribution and physical conditions of the intervening ionised medium across the Galaxy \citep{Abadi2024}.
    
    \item 
    Absorption in the ionised surroundings of the SNR: Low-frequency observations are a crucial tool for tracing the ionised interfaces formed where SNR shock fronts interact with adjacent molecular clouds (e.g.,  \citealt{Brogan2005}, \citealt{Castelletti2011}, \citealt{Castelletti2021}). They can also reveal ionised shells shaped by the radiation of the supernova progenitor, whose imprint remains in the surrounding material \citep{Arias2019c}. 
    \item 
    Intrinsic absorption: SNRs can contain reservoirs of ionised gas capable of absorbing the synchrotron radiation they generate, an effect that becomes clear at low ($<100$~MHz) radio frequencies. This absorption can originate from unshocked ejecta-material interior to the reverse shock that has adiabatically cooled since the explosion (e.g., \citealt{Kassim1995}, \citealt{Arias2018} for Cas~A).

\end{itemize}

From the radio perspective, thermal free-free absorption inferred from SNR spectra can be parameterised through a model that, assuming a fixed electron temperature, provides an estimate of the emission measure (EM; \citealt{Abadi2024}). However, this parameterisation is inherently degenerate: different combinations of electron density and path length can yield similar EM values, and therefore similar turnover frequencies. This degeneracy makes it difficult to uniquely constrain the physical nature and location of the absorbing material using radio continuum data alone, thus requiring complementary multiwavelength observations.
In the case of low-frequency turnovers produced by ionised gas unrelated to the SNR, extrinsic free-free absorption may arise in H\,{\sc ii} regions (or their envelopes) or in diffuse ionised gas located either along the line of sight or in the periphery of the remnant. Since radio spectra alone do not provide information on the relative location of the emitter and absorber, complementary observations are required. In particular, infrared data tracing typical emission from H\,{\sc ii} regions (e.g. at 8-24~$\mu$m) and existing catalogues of ionised regions can be used to assess whether the absorbing material is foreground to the remnant. Distinguishing this scenario from absorption intrinsic to the remnant or arising at interaction sites requires spatial comparison with tracers at other wavelengths.
In the case of free-free absorption arising at the interface between the SNR and a molecular cloud, multiwavelength data provide key diagnostics. For instance, in the interacting SNR 3C 391, X-ray data indicate enhanced absorption of soft emission towards regions of higher column density, while the key evidence comes from the spatial correspondence between the radio absorption, molecular gas traced by CO emission, and ionised fine-structure atomic lines in the 12–18 $\mu$m range \citep{Brogan2005}. In IC 443, near-infrared [FeII] emission spatially correlated with free-free absorption, as revealed by VLA observations at 74 MHz, has been used to interpret the low-frequency radio absorption in terms of  shocked gas at the interaction interface \citep{Castelletti2011}. Other examples of such a scenario, including W49B, Kes 73, and 3C 391, further support this interpretation \citep{Castelletti2021}.
In relation to intrinsic free-free absorption arising from cold, unshocked ejecta, a notable example is Cas A \citep{Arias2018}. This scenario, in which the absorbing material is located within the remnant interior, is observationally rare and has so far been identified only in a few young SNRs. In Cas A, low-frequency absorption correlates with infrared emission tracing unshocked ejecta, indicating that both diagnostics arise from the same cold material internal to the shell. In some cases, however, low-frequency departures from a power-law spectrum may also reflect shock acceleration signatures, indicative of multiple electron populations, or genuine curvature in the spectrum (\citealt{Anderson1993}, \citealt{Castelletti2007}, \citealt{Reynoso2015},  \citealt{Arias2019a}). 

Extending the analysis towards the higher-frequency regime, which can be probed by the SKA, the spectral behaviour above about 10~GHz remains poorly studied due to observational limitations.   Radio interferometers filter out large spatial scales, severely hampering accurate flux density measurements, and the problem worsens as frequency increases (see Section~\ref{sec:sd}). 
Conversely, single-dish observations at these frequencies suffer from limited spatial resolution.
Besides these technical difficulties, it is important to note that the intensity of the synchrotron emission 
decreases significantly with frequency, making SNRs fainter.
However, there are at least three main reasons for observing in this part of the spectrum: 
(i) 
assuming that synchrotron emission remains the dominant component at these frequencies, data above 10~GHz allow for a better determination of $\alpha$, whose uncertainty scales as 
$\Delta \alpha\propto{1}/{\left|\ln({\nu_1}/{\nu_2})\right|}$;
(ii) 
variations of $\alpha$ may occur \citep[concave-up or concave-down spectra, see][]{Urosevic2014,2016A&A...586A.134P,2019MNRAS.482.3857L}, whose origin is worth investigating;  (iii) Faraday rotation is less severe than at lower frequencies, resulting in lower depolarisation (see Section~\ref{sec:pol}).

It is worth emphasising the importance of spectral index maps \citep[e.g.,][]{Loru2024} for studying the electron energy distributions in the SNR interior. The imaging capabilities of the SKA, combined with its five frequency bands covering a wide frequency range, will allow detailed 
mapping of the radio spectral index across different frequencies. 
This will provide valuable constraints on spatial variations in the electron spectrum across an SNR, thereby improving our understanding of particle acceleration at the shock and the subsequent downstream evolution.

\subsection{Interaction with molecular clouds}
As an SNR expands, the hot plasma pushes and compresses 
the atomic and molecular gas in contact with the remnant \citep{Chevalier1974},
injecting energy, mass, and momentum into the ISM and profoundly affecting its physical and chemical properties \citep{Slane2016}. 
On typical cloud scales ($<10$~pc),
SNRs drive slow shock compression that impact 
nearby pre-existent molecular clouds (MCs), 
locally enhancing the density of the molecular material \citep{Cosentino2019,dellova2020, ustamujic2021}. 

The interaction between SNRs and the surrounding ISM  is a complex, time-dependent process with important astrophysical implications that are not yet fully understood. Firstly, SNR shock waves are believed to be the primary mechanism for accelerating charged particles to relativistic energies. Recent observations suggest that SNR-MC interactions create favourable conditions for such acceleration, leading to the production of CRs with energies of up to several PeV \citep{Bykov2018, Tatischeff2018}. When the SNR shock encounters the dense material of MC, it rapidly decelerates. However, the downstream plasma does not respond immediately; it continues to expand at the velocity it had prior to the SNR–MC interaction. Under this configuration, the effective compression ratio experienced by high-momentum cosmic rays increases \citep{2026A&A...709A.175P}. As a result, the acceleration efficiency of the highest-energy cosmic rays is enhanced.
Secondly, SNR-MC interactions play a dual role in the formation of new stars. On one hand, the compression and cooling of MCs by SNR shocks can trigger star formation \citep{Herbst1977, Inutsuka2015, Klessen2016}, inducing the gravitational collapse of the densest clumps. Although newborn stars are expected to condense out of the expanding shell, conclusive observational evidence for this phenomenon remains scarce \citep{2015A&ARv..23....3D}. On the other hand, the same interaction can enhance turbulence within the cloud, dispersing the material and suppressing further star formation. Despite their importance, the observational constraints on supernova feedback effects on the ISM remain limited. While SNR-MC interactions have been shown to enhance the star formation efficiency in nearby galaxies by up to $>40\%$ \citep{Rico-Villas2020}, the first and only direct evidence of SNR-triggered star formation in a dense cloud has been reported only recently 
\citep{Cosentino2025}.

The search for SNR-MC interactions 
mainly relies on morpho-kinematic indicators in the molecular material. The shock front injects momentum 
into the ambient medium, accelerating it to a small fraction of the shock velocity. 
This acceleration appears as a broadening of molecular line profiles, with FWHMs of several tens of km~s$^{-1}$ \citep{Kilpatrick2016,Rho2017, Rho2021}, compared with the narrow line widths ($<5$~km~s$^{-1}$) typical of quiescent, undisturbed clouds. 
Likewise, column density enhancements or sudden jumps in the velocity distribution of the shocked gas serve 
to pinpoint the interaction regions \citep{Cosentino2019}. 

One of the most widely used molecular tracers of 
SNR-MC interaction is CO, 
the second most abundant molecule in the ISM after H$_{2}$. Its low-$J$ rotational 
transitions are the most common 
means of mapping the distribution of the cold ambient gas 
that may be impacted by an SNR blast wave, as its 
high abundance makes this tracer relatively easy to detected. However, establishing 
an association between molecular clouds and SNRs 
based solely on CO observations is 
challenging, as line-of-sight confusion 
and superposition 
are frequent in the Galactic plane, and the 
distances to most SNRs remain uncertain. 
This highlights the need for additional, more specific shock tracers that provide unambiguous evidence of interaction, such as OH and methanol (CH$_{3}$OH) masers, which require the stringent excitation conditions found in shocked regions (see Section~\ref{sec:mas}). 
Silicon monoxide (SiO) is also a well-known molecular tracer of shocks \citep{Schilke1997}. While highly depleted in quiescent regions 
(abundances relative to H$_2$ $< 10^{12}$; \citealt{MartinPintado1992}), its abundance can be enhanced by up to a factor of $10^6$ in shocked gas 
\citep{JimenezSerra2005}, as the passage of the shock leads to dust-grain destruction through sputtering and erosion, realising Si into the gas phase and promoting SiO formation.

Considering these different components, 
synergistic multi-wavelength studies are 
key to fully understanding the physical conditions 
of SNRs and their interactions with nearby MCs. 
In particular, radio continuum observations, especially spatially resolved spectral-index analyses, can identify regions of non-thermal emission that likely trace shocks, while (sub)-millimetre molecular spectroscopy can constrain the physical parameters of these regions, such as density, temperature, and kinematics. 
Nevertheless, such comprehensive 
analyses have so far been performed only for a handful of selected objects \citep[e.g.,][]{Mazumdar2022,Zhong2023,Loru2025}. 
Recent large-area molecular line surveys of the Milky Way, such as MOPRA, FUGIN (both targeting CO and its isotopologues),  and SEDIGISM (including CO and other species such as 
CH$_{3}$OH, H$_{2}$CO, and HC$_{3}$N),
now provide an opportunity to extend such studies to a much larger fraction of the Galactic SNR population. Furthermore, 
joint projects with existing (sub-)millimetre  facilities such as NOEMA, ALMA, and its upcoming WSU development, as well as future instruments like AtLAST, offer great synergies with the unique capabilities of SKA. Together, these  
multi-wavelength 
approaches will 
enable a far more complete view of the interplay between SNRs and the ISM.

Finally, there is growing evidence that our own Sun might have formed within 
a proto-solar nebula impacted by at least one SNR \citep{Boss2013,Young2014,Parker2023}. Therefore, achieving a 
deeper understanding of SNR-MC interactions will directly 
inform our knowledge of the origin of the Solar System and, ultimately, of life on Earth.

\subsection{OH masers in SNRs}
\label{sec:mas}

Current studies of SNR kinematics primarily rely on radial-velocity measurements and numerical simulations. While radial velocities provide kinematic information along the line of sight, 
they do not fully constrain the 3D motion of SNRs or accurately describe their dynamical 
evolution. 
Although numerical simulations can model complete 3D kinematics, they often rely on theoretical assumptions that may deviate from actual physical conditions, making observational validation essential. Fundamental questions about SNR kinematics remain unresolved, for example about a reliable estimation of the characteristic expansion velocities of shocked shells or the predominant direction of the shocked gas (perpendicular to or tangential to the shell structure). Direct 3D kinematic observations would provide critical insights into SNR dynamics and their interaction with the surrounding 
ISM. Moreover, such observations would enable empirical tests and subsequent refinements of SNR dynamical evolutionary models.

Distance is an essential parameter for determining SNR 3D kinematics and deriving key physical properties, including size, luminosity, and age. Various approaches have been adopted to measure the distances 
to Galactic SNRs, including kinematic methods, the $\Sigma$--$D$ (surface brightness--physical diameter) relationship, optical extinction methods, and 
estimates based on associated objects \citep{Milne+1970,Green+1984,Wang+2020}. However, significant discrepancies often exist between distance estimates obtained 
using different methods, as each technique depends on 
particular physical assumptions or is subject to substantial systematic uncertainties. Trigonometric parallax measurements are free from physical assumptions and provide a valuable means to address 
these inconsistencies in SNR distance estimates.

The 1720~MHz~OH maser serves as a unique probe for studying the distances and 3D kinematics of SNRs, 
as it is recognised as a signpost for 
SNR-MC interaction \citep{Lockett+1999} and, being compact, can persist for decades.  To date, 1720~MHz~OH masers have been found in 
approximately 25 SNRs \citep[e.g.,][]{Frail+1996,Green+1997,Hewitt+2008}. Four of 
these (W28, W44, W51C, and IC~443) have been observed via phase-referencing observations using the Very Long Baseline Array (VLBA) in the early 2000s \citep[e.g.,][]{Hoffman+2005a,Hoffman+2005b}. 
Several SNRs with detected 1720 MHz OH maser emission--including G349.7+0.2, GBT~37A, 3C~391, W44, IC~443, and W51C \citep{Brogan+2013}--show simultaneous $\gamma$-ray emission. This correlation suggests that OH masers may trace key regions of 
CR acceleration, expanding their scientific significance from traditional shock diagnostics to high-energy astrophysics.
Shocks in maser-emitting regions are generally classified as either J-type (jump) or C-type (continuous). J-type shocks are dissociative and occur for shock velocities $v_{s} > 30-50~{\rm km~s^{-1}}$, whereas C-type shocks are non-dissociative and typically occur at  $v_{s} < 40-50~{\rm km~s^{-1}}$. 
Nevertheless, the properties of SNR 1720~MHz~OH masers remain poorly characterised.

\begin{figure}
    \centering
    \includegraphics[width=\linewidth]{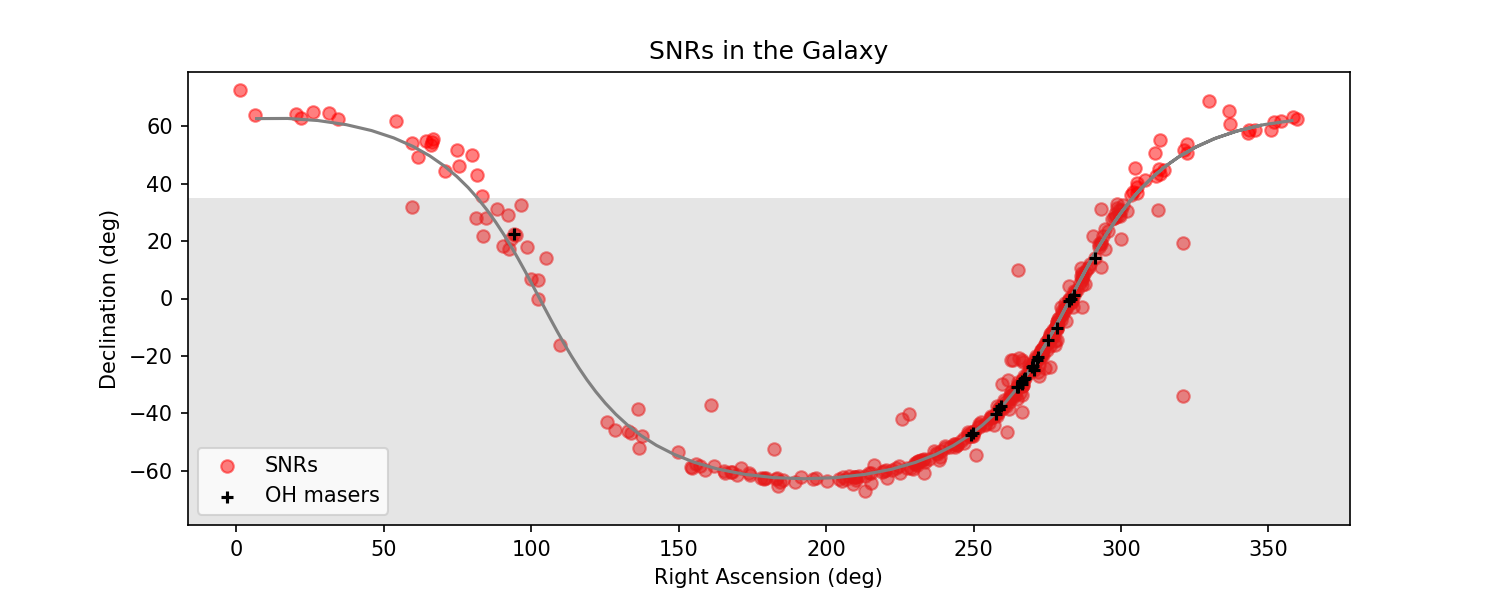}
    \caption{Distribution of SNRs in the Milky Way. Red dots mark the positions of known SNRs \citep{Ferrand+2012}, while black crosses indicate those associated with 1720-MHz OH emission.  
    The grey solid line traces the Galactic disc, and the shaded region highlights the portion of the sky observable with SKA-VLBI ($\delta < 35^{\circ}$; \citealt{Li+2024}).}
    \label{fig:SNRs_OH_masers}
\end{figure}

The detected SNRs are predominantly concentrated in the Galactic centre region (see Figure~\ref{fig:SNRs_OH_masers}). This spatial distribution results in suboptimal 
$uv$-coverage for current very long baseline interferometry (VLBI) arrays, most of which are located in the northern hemisphere. Poor 
$uv$-coverage can distort the synthesised beam, leading to an overestimation of the apparent sizes of maser spots. Additionally, due to the limited sensitivity of current VLBI arrays, typically only a few maser spots are detected per SNR, further reducing astrometric precision. These observational hurdles, compounded by the inherent difficulties of low-frequency ($< 8$~GHz) astrometry, make high-precision 
measurements particularly demanding. As a result, no proper motions or parallaxes have been 
successfully measured for 1720~MHz~OH masers associated with SNRs.

SKA-VLBI will enable 3D kinematic studies of SNRs through precise trigonometric parallax and proper motion measurements of their associated 1720~MHz~OH masers. As a southern-hemisphere facility, SKA-VLBI will offer extensive coverage of the majority of Galactic SNRs (see Figure~\ref{fig:SNRs_OH_masers}). With its unprecedented sensitivity, SKA-VLBI will not only detect more maser spots and fainter components per SNR but also facilitate the discovery of new SNRs harbouring such maser emission.

The enhanced sensitivity will enable the detection of more 1720~MHz~OH masers, facilitating detailed studies of their physical properties (see Section \ref{sec:calc}).
SKA-VLBI will provide transformative multi-scale observational capabilities, achieving angular resolutions ranging from $\sim$250~mas down to $\sim$4~mas (corresponding to baselines up to 10000~km at 1720~MHz; see \citealt{Li+2024}). 
This unprecedented resolution will 
allow systematic investigation of morphological variations in maser-emitting regions, which is critical for determining whether observed structures represent true physical scales or unresolved blends.

For 1720~MHz~OH masers in Galactic SNRs, the $\sim$4~mas resolution corresponds to a remarkable 
physical scale of $\sim$20~AU at a distance of 5~kpc. 
This 
precision is sufficient to definitively test the compact core hypothesis ($\sim$50~AU scale) suggested by \citet{Hoffman+2005a}. High-resolution ($\sim$4~mas) observations will determine whether individual maser spots consist of 
single compact components or multiple substructures, thereby constraining the physical size and geometric morphology (e.g., disk-like, jet-like, or shell-like configurations) of the emission region. 
At this resolution, SKA-VLBI can also distinguish the shape of 
maser emission regions at distances up to 12.5~kpc, if the compact core of 1720~MHz~OH masers indeed follows the $\sim$50~AU scale 
proposed by \citet{Hoffman+2005a}.

SKA-VLBI can map the flux distribution of maser spots. The morphology of 1720~MHz~OH masers in SNRs is closely linked to the 
pre- and post-shock structures. High-resolution observations can distinguish shock-compressed regions (narrow linewidths) from turbulent regions (dispersed structures), thus providing crucial tests of shock propagation models.

According to theoretical collisional pump models, 1720~MHz~OH masers in SNRs arise under limited physical conditions, requiring a moderate temperature range of $T = 50 - 125~{\rm K}$, a density of $n \sim 10^{4} - 10^{5}~{\rm cm^{-3}}$, and an OH column density of $N_{\rm OH} \sim 10^{16}~{\rm cm^{-2}}$ \citep{Lockett+1999}. 
By combining these observations with complementary atomic and molecular gas data, 
it will be possible to systematically probe the OH excitation environment, thereby placing strong constraints on the physical conditions required for SNR 1720~MHz~OH maser emission.

The MultiView technique is ideally suited for astrometry of 1720~MHz OH masers (see Section \ref{sec:calc}). 
Given that 
such masers can 
persist for decades, their proper motions can be derived by comparing 
spot positions across multiple epochs. 
By jointly modelling their proper motions and radial velocities, it becomes possible to determine accurate distances and reconstruct their full 3D motions. 
The resulting 3D kinematics provide a direct means of measuring the expansion velocities of associated shock fronts. 

For parallax measurements, phase referencing is fundamentally limited to $\sim$0.70~mas precision, restricting reliable distance determinations to SNRs 
within $\sim$140~pc while maintaining 10\% precision. In contrast, the MultiView technique with $\sim$3000~km SKA-VLBI baselines, achieves parallax precision of $\sim$40~$\mu$as, enabling 10\%-accurate distance measurements out to $\sim$2.5~kpc. This order-of-magnitude improvement facilitates a survey of all nearby SNRs and 
help resolve long-standing distance controversies that have limited detailed studies of remnant kinematics and evolutionary processes.


\subsection{Radio to \texorpdfstring{$\boldsymbol{\gamma}$}{gamma}-ray characterisation}
\label{sec:gamma}
A central open question in high-energy astrophysics concerns the role of SNRs in the origin of Galactic CRs.
The expanding shock fronts of SNRs have long been recognised as prime sites for particle acceleration up to energies of at least a few hundred TeV (\citealt{Blandford1978}; \citealt{Reynolds2008}). Whether these shocks can truly operate in the Pevatron regime remains an open issue, and other potential contributors--such as pulsar wind nebulae embedded within composite remnants or superbubbles acting over larger spatial scales--may also play a significant role in shaping the Galactic CR population \citep{Cristofari2021}.

Numerous studies have established the leading role of SNRs as efficient accelerators of high-energy particles \citep{Ackermann2013}, particularly in scenarios involving the re-acceleration of pre-existing CRs and their compression within crushed clouds in SNR-MC interaction systems (\citealt{Uchiyama2010}; \citealt{Lee2015}; \citealt{Tang2015}; \citealt{Cardillo2016}). Nevertheless, several key questions remain regarding the physical mechanisms governing particle acceleration. In particular, direct evidence for proton acceleration is still debated  due to the difficulty of disentangling the $\gamma$-ray emission components produced by leptonic processes (e.g., inverse Compton scattering and bremsstrahlung) from those arising from hadronic interactions, such as the decay of neutral pions generated by CR collisions with ambient matter \citep{Cardillo2016}. Moreover, determining the maximum particle energy achievable in SNRs--and its evolution over time--continues to be a central challenge in shock-acceleration theory (\citealt{Celli2019}; \citealt{Morlino2021}).

Addressing these questions requires the reconstruction of broadband SEDs of SNRs, spanning radio to $\gamma$-rays energies. Radio observations play a key role, providing the most direct evidence of relativistic electrons through synchrotron emission. Anchoring the radio portion of the SED ensures physical consistency across the spectrum and allows constraints on particle injection and acceleration efficiencies, magnetic field strengths, and absorption effects in the surrounding medium. These parameters directly influence the predicted $\gamma$-ray emission, whether through leptonic or hadronic channels. Without the radio anchor, broadband SED modelling tends to extrapolate from $\gamma$-ray data towards lower energies, potentially leading to ambiguous interpretations of the particle energy distribution and the dominant emission mechanisms.

Spatially resolved studies of the radio spectral index provide further insight into the local particle-acceleration processes and the physical conditions within SNRs. Large and morphologically complex remnants--typically exceeding 
20~arcmin in size--are particularly well suited to such analyses, as their extent and structural diversity allow the identification of regions under different environmental conditions (see e.g.,  \citealt{Lee_2008}; \citealt{Uyaniker_2004}; \citealt{Egron2017}). Variations in shock strength, ambient density, and interactions with atomic or molecular clouds generate distinct populations of relativistic electrons, each characterised by its own energy distribution and non-thermal spectrum. Mapping the spectral index across a remnant can pinpoint these populations, while co-spatial bright, flat-spectrum radio and $\gamma$-ray emission often highlights localised regions of efficient, potentially hadronic, CR acceleration. This approach has been successfully applied to several extended $\gamma$-ray-emitting SNRs such as Vela \citep{Alvarez2001}, W44 \citep{Castelletti2007}, S147 \citep{Xiao2008}, IC~443 \citep{Castelletti2011}, and the Cygnus Loop (\citealt{Loru2021}; Castelletti et al. \textit{in prep}.). These remnants exemplify how spatially resolved radio data enables the study of shock-cloud interactions and local CR acceleration, complementing the integrated SED view and providing tighter constraints on theoretical models describing CR energetics.

Even more compact remnants can offer key insight when multi-wavelength data are available. For instance, \citet{Loru2025} investigated Kes~73 through combined radio, CO-line, and $\gamma$-ray observations, revealing a shock-MC interaction region where the radio spectral index was used to model the broadband SED from radio to $\gamma$-rays. This study supported a lepto-hadronic origin for the emission, constrained the maximum and total energies of the accelerated particles, and provided valuable information about the local environment, including magnetic field strengths and ionised gas densities.

The prospects for such analyses are rapidly expanding. SKA precursors, with their combination of high sensitivity and ability to recover extended spatial scales, now enable spectral index mapping for a broader sample of SNRs. \citet{Loru2024} leveraged data from GLEAM \citep{Wayth_2015} and the SMGPS \citep{Goedhart_2024} to produce spectral index maps at $\sim\!2^{\prime}.5$ resolution for 29 known SNRs. These maps have proven useful for constraining SNR morphologies, tracing low-brightness diffuse emission, distinguishing regions affected by different shock and environmental conditions, and identifying superimposed sources such as H\,{\sc ii} 
regions that can bias spectral index determinations.

At $\gamma$-ray energies, ongoing and forthcoming surveys promise to complement these radio efforts. The \textit{Fermi}-LAT mission continues to provide all-sky coverage with improved sensitivity to extended Galactic sources, enabling detailed studies of their $\gamma$-ray morphology \citep{2025A&A...701A.206G}, while the Large High ALtitude Air Shower Observatory (LHAASO; \red{\citealt{Cao2021}}) has recently revealed PeV $\gamma$-ray emission from several SNR candidates, shedding new light on particle acceleration at the highest energies. In the near future, the Cherenkov Telescope Array Observatory (CTAO; \citealt{Hofmann2023}; Section~\ref{sec:cta})
will deliver unprecedented angular resolution and sensitivity across the TeV domain, enabling detailed morphological and spectral studies of SNRs on sub-arcminute scales. Together with next-generation radio facilities, these developments make the comprehensive reconstruction of broadband SEDs--from radio to PeV energies--an increasingly attainable goal. The interaction of supernova remnants and dense ambient material also provides an ideal environment for multi-messenger studies. This is because molecular clouds "illuminated" by cosmic rays accelerated at the shock front are expected to produce neutrino emission \citep{2006APh....26..310V}. In this framework, SKA will then provide important constraints for neutrino telescopes (as KM3Net, see Section \ref{sec:km3net}).

\subsection{Polarimetry and turbulence}
\label{sec:pol}

Polarimetry of the synchrotron emission is a unique tool to directly probe the spatial and spectral structure of the magnetic field. 
We are just at the beginning of a deep exploration of the polarisation images of SNRs. 
Indeed, while polarisation maps of SNRs have been known since the early observations in the mid-20th century \citep{2015A&ARv..23....3D}, the methodology for theoretical synthesis of such images from the three-dimensional (3D) magnetohydrodynamic (MHD) simulations has only been developed recently \citep{2015MNRAS.449...88S,2017MNRAS.466.4851V,2017MNRAS.470.1156P,2023MNRAS.518.6377P}. 

The SKA is expected to produce high-resolution maps of the Stokes parameters. 
By 
comparing them with images synthesised from 
detailed 3D MHD models, one may infer important conclusions about the present-day state of the magnetic field and relativistic electrons in an SNR--namely, their spatial distributions and energy spectra. Moreover, it will be possible to place constraints on the properties of the 
supernova progenitor star from the polarisation pattern of the remnant. For example,  
\citet{2023MNRAS.518.6377P} 
synthesised polarisation images from a 3D numerical model of SN~1987A and demonstrated that the progenitor of 
this supernova was a very slow rotator. Hopefully, the possibilities offered by coupling the 3D modelling 
with high-resolution polarisation imaging will 
help resolve 
long-lasting questions, e.g.,  why 
radial magnetic polarisation is mostly detected in young SNRs, while 
tangential polarisation dominates in evolved ones \citep{2015A&ARv..23....3D}. 

Another important piece of information comes from the polarisation fraction $\Pi$. Typical values for SNRs 
range between $\sim5\%\div 25\%$, 
which is well below the theoretical limit $\approx 70\%$. Such low values indicate significant randomness in the magnetic field. The measured Stokes parameters are 
integrated along the line of sight 
within the SNR, where both the strength and orientation of the magnetic field component vary. 
An internal Faraday effect, which is also sensitive to the density structure, 
contributes to the depolarisation of 
the signal. 
However, projection effects alone could lower $\Pi$ to values around $20\%$, as shown by detailed 3D modelling \citep{2023MNRAS.518.6377P}, if only the ordered magnetic field component is considered. 
Polarisation fractions below such a value are due to the turbulent component of the magnetic field. 
The synchrotron emission theory generalised to electron emission in a magnetic field with ordered $B$ and disordered $\delta B$ components \citep{2016MNRAS.459..178B,2024A&A...689A.137B} demonstrates that $\Pi$ could be as low as $10\%$ solely due to turbulence if the ratio $\delta B/B$ 
in the plane of the sky is about $3$. The spatial variation of polarisation fraction over the SNR surface highlights differences in the properties of turbulence and 
provides insights into the plasma microphysics operating in SNRs. 
For example, \citet{2024ApJ...973..105S} 
used such variations to confirm different regimes of electron acceleration in Kepler's SNR. In particular, it has been shown that a high level of turbulence increases the efficiency of electron acceleration, causing a shortening of the acceleration 
timescale in regions characterised by a low polarisation fraction \citep{2022ApJ...935..152S,2024ApJ...973..105S}.

The SKA will also create a synergy with polarisation measurements in the X-ray band. 
Indeed, the Imaging X-ray Polarimetry Explorer (IXPE; Section~\ref{sec:ixpe})
has detected 
polarised synchrotron emission from a number of SNRs \citep{2024Galax..12...59S}. 
For instance, the joint use of 
radio and X-ray polarisation measurements 
confirms the existence of a prominent turbulent component with a strength $\delta B\sim B$ in SN~1006 and Tycho's SNR \citep{2024A&A...689A.137B}. 

Given such a high level of disordered field in SNRs, 
observations of these objects may be used to reveal the properties of MHD turbulence. 
Indeed, \citet{2018MNRAS.480.2200S} 
discovered from 
radio images that the turbulence in the remnant of Tycho's supernova is of the Kolmogorov type, i.e., the power spectrum of magnetic turbulence follows the $k^{-5/3}$ law  
over a one-dimensional contours.
\citet{2025arXiv250923295P} 
performed a two-dimensional autocorrelation analysis of the radio and X-ray images of Tycho's SNR and confirmed the findings of \citet{2018MNRAS.480.2200S} by demonstrating that the power spectrum of density and magnetic field fluctuations follows the two-dimensional law $k^{-8/3}$. 
At this point, it merits emphasising the importance of SKA's high-resolution polarisation imaging capabilities. 
These will allow one to study the properties of turbulence over a wide range of length scales, from the SNR size (a fraction of a degree) 
down to sub-arcsec resolution ($\sim0^{\prime\prime}.2$ at 1.3~GHz in the AA4 configuration), 
i.e., over a few orders of magnitude. Interestingly, with a resolution $0^{\prime\prime}.2$, one could eventually resolve the fastest-growing mode 
of the \citet{2004MNRAS.353..550B} instability in an SNR at a distance $d\lesssim 3.5\,B_{3}E_{2}/(V_{4}^{3}n)\ \mathrm{kpc}$, where the ISM magnetic field is in 3~$\mu$G, density $n$ in cm$^{-3}$, maximum energy of CRs $E$ in $10^2$~TeV, and the shock speed $V$ in $10^4$~km~s$^{-1}$.

\subsection{Insights from MHD modelling}

The observational progress discussed in previous sections sets the stage for studying SNRs as natural laboratories for shock physics, magnetic fields, and particle acceleration. With the SKA’s superior resolution, sensitivity, and polarisation capabilities, these processes can now be investigated in unprecedented detail, enabling breakthroughs in understanding SNR microphysics and evolution. 
To fully exploit this potential, theoretical frameworks capable of capturing the non-linear interactions in SNRs are essential. 
MHD modelling--particularly in 3D and coupled with particle acceleration prescriptions--has become a cornerstone 
for interpreting SNR observations and investigating the origin of non-thermal emission and the properties of particle acceleration at 
shock fronts (e.g.,  \citealt{2007A&A...470..927O, 2012ApJ...749..156O}). The integration of MHD simulations with SKA observations will not only deepen our understanding of individual remnants but will also enable statistically 
robust population studies that trace the diversity of stellar death.

MHD simulations provide a self-consistent treatment of the evolution of 
plasma, magnetic fields, and shocks that define the morphology, structure and emission of SNRs. More specifically, they allow us to model: (i) shock dynamics and instabilities (e.g., Rayleigh–Taylor, Kelvin–Helmholtz, Richtmyer–Meshkov; e.g.,  \citealt{1992ApJ...392..118C}); (ii) anisotropic expansion shaped by asymmetries in the explosion (e.g.,   \citealt{2021A&A...645A..66O}) or in the surrounding CSM (\citealt{2024ApJ...977..118O}); 
(iii) polarised synchrotron emission, crucial for mapping magnetic field geometry (e.g.,  \citealt{2023MNRAS.518.6377P}); and (iv) particle acceleration feedback, using a spatially and temporally varying effective adiabatic index (e.g., \citealt{2010A&A...509L..10F, 2012ApJ...749..156O, 2017MNRAS.468.1616P, 2018ApJ...852...84P}) that accounts for compression, turbulence, and CRs streaming effects. By generating synthetic observables (such as radio intensity maps, polarisation vectors, and Faraday rotation measures; e.g.,  \citealt{2007A&A...470..927O, 2011A&A...531A.129B, 2023MNRAS.518.6377P}), MHD simulations enable direct comparison with observations. These comparisons are crucial for disentangling the roles of explosion physics, environmental structure, and magnetic processes in shaping the observed characteristics of SNRs.

SKA’s capabilities align directly with the outputs of modern MHD models. 
Indeed, SKA is expected to resolve filamentary structures and fine-scale shock features in young and middle-aged SNRs 
nearby (e.g.,  SN~1987A, G292.0+1.8). Theoretical studies predict that these features are closely 
linked to local magnetic field amplification and turbulence, phenomena that shape the local synchrotron emissivity and polarisation. In addition, the wide frequency coverage enables detailed spectral index mapping, which in turn constrains the 
spatial and energetic distributions of accelerated particles. Coupling this with MHD models that include evolving magnetic-field strengths allows us to infer local shock obliquity and test non-linear 
DSA models. 

SKA is expected to produce full-Stokes imaging, 
mapping the polarisation of radio synchrotron emission across the entire remnants. MHD models predict distinct 
polarisation signatures for radial versus tangential magnetic fields, as well as for regions dominated by turbulence. Simulations can produce synthetic Stokes maps (e.g.,  \citealt{2017MNRAS.470.1156P,2023MNRAS.518.6377P}), providing 
benchmark for interpreting SKA observations. Through Faraday rotation and depolarisation studies, SKA can also probe the line-of-sight component of magnetic fields and thermal electron distributions. MHD simulations, especially those incorporating realistic circumstellar structures (e.g.,  wind-blown bubbles, clumps), are essential for reconstructing 3D magnetic and density structures (e.g.,  \citealt{2019A&A...622A..73O}).

The combination of MHD modelling and SKA observations will unlock transformative science cases for the study of SNRs. By directly comparing synthetic morphologies and polarisation maps from simulations with SKA’s high-resolution radio data, researchers will be able to place tight constraints on explosion asymmetries and on the structure of the circumstellar and interstellar media shaped by progenitor stars. Young, bright remnants such as G292.0+1.8 and SN~1987A are prime targets for such investigations, as they preserve signatures of both the explosion geometry and the interaction between rapidly expanding ejecta and a non-uniform environment. These comparisons are key to determining the nature of the progenitor (discriminating between red supergiants, Wolf-Rayet stars, or white dwarfs) and to advancing our understanding of stellar death pathways, explosion physics, and early remnant evolution.

In particular, SN~1987A provides a compelling case study, as it is evolving within a highly structured 
CSM sculpted by the progenitor’s complex mass-loss history. MHD models have shown that this interaction significantly shapes the remnant’s morphology, magnetic field topology, and non-thermal emission (e.g.,  \citealt{2019A&A...622A..73O}). These simulations have successfully reproduced key observational features, such as the toroidal structure and asymmetric brightness distribution seen in synchrotron radiation, underscoring the importance of environmental anisotropies. In more evolved systems such as the Vela SNR, Puppis~A, and 
remnants in the Large Magellanic Cloud (LMC), like N63A, N49, and SNR~0540$-$69.3, expansion into the 
ISM leads to interactions with large-scale magnetic fields and ambient density fluctuations, producing intricate morphologies and complex polarisation patterns. These features provide a window into the long-term evolution of shock fronts and the role of the ISM in shaping remnant dynamics.

SKA’s unmatched sensitivity, angular resolution, and polarisation capabilities will make it possible to map magnetic-field orientations, degrees of order, and spectral-index variations across SNRs with unprecedented precision. These observational diagnostics can be directly compared with synthetic observables from MHD models to investigate the role of magnetic-field amplification, turbulence, and shock geometry in shaping the observed emission. In the case of SN~1987A, such comparisons will offer critical insight into how core-collapse SNe evolve in non-uniform environments, while for older remnants they will illuminate how expanding shocks interact with Galactic magnetic fields on larger scales. This synergy between observations and modelling will provide a coherent, physically grounded framework to interpret the morphological and spectral diversity of SNRs at various evolutionary stages.

Moreover, the amplification of magnetic fields at SNR shocks is central to understanding both synchrotron emission intensity and the efficiency of 
CR acceleration. Theoretical work predicts that magnetic fields can be enhanced by orders of magnitude (reaching hundreds of microgauss) through mechanisms such as shock compression, turbulent cascades, and 
CR streaming instabilities. SKA’s fine spatial resolution will enable the direct imaging of these amplified structures, revealing the scales and conditions under which they arise. When paired with MHD simulations that include feedback from accelerated particles (e.g.,  \citealt{2012ApJ...749..156O, 2018ApJ...852...84P}), these observations will help unravel the complex interplay between fluid dynamics, magnetic-field amplification, and CR production.

Finally,
embedding acceleration physics (such as 
DSA) within MHD simulations enables the generation of testable predictions that can be directly compared with SKA data (see Sections \ref{sec:sed} and \ref{sec:gamma}). This approach will allow researchers to evaluate and refine competing models for how particles gain energy in shocked plasmas, advancing our theoretical understanding of CR 
origins and SNR energetics.


\section{Technical and observational feasibility}
\label{sec:tech}
In the previous sections, we have outlined several science cases that will benefit from the SKA. We now turn to a quantitative evaluation of the instrumental and observational feasibility of these studies, addressing both instrumental requirements and the strategies needed to recover large-scale structures.

\subsection{Technical requirements}
\label{sec:calc}

\begin{figure}
    \centering
    \includegraphics[height=8cm]{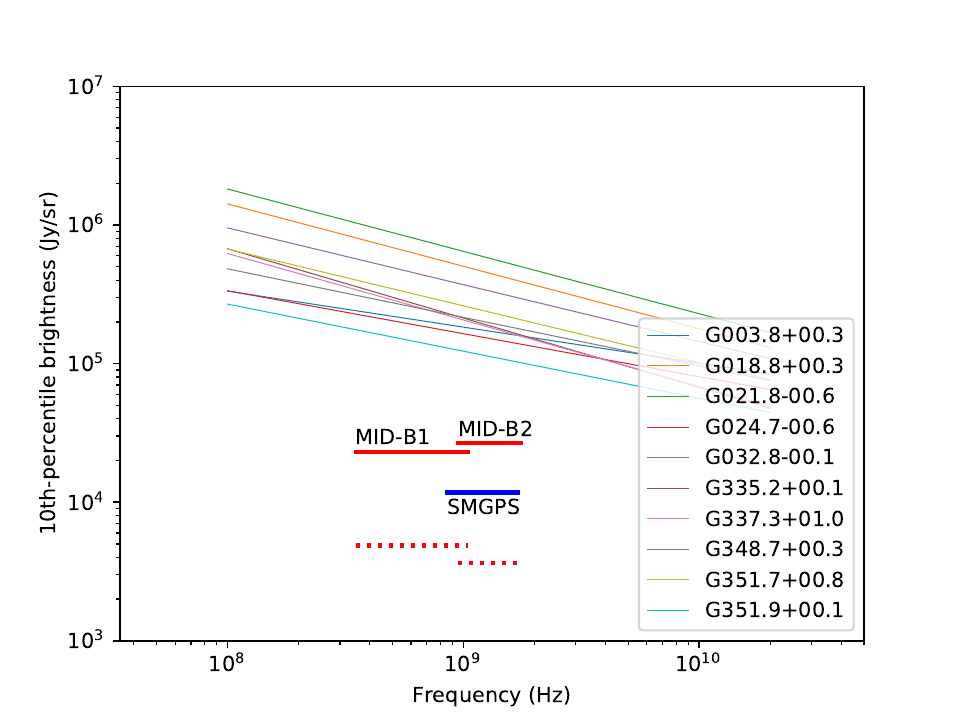}
    \caption{Simulated SEDs of the 10 SNRs in our sample (narrow lines). 
    The sensitivity limit and frequency 
    coverage of the SMGPS are shown as a thick blue line. 
    The solid red lines indicate the sensitivity limits achieved by SKA-MID in Bands 1 and 2 for a one-hour integration time, while the dashed red lines show the corresponding limits when the SKA images are convolved to match the SMGPS beam size. }
    \label{fig:sens_mk}
\end{figure}

\begin{figure}
    \centering
    \includegraphics[height=8cm]{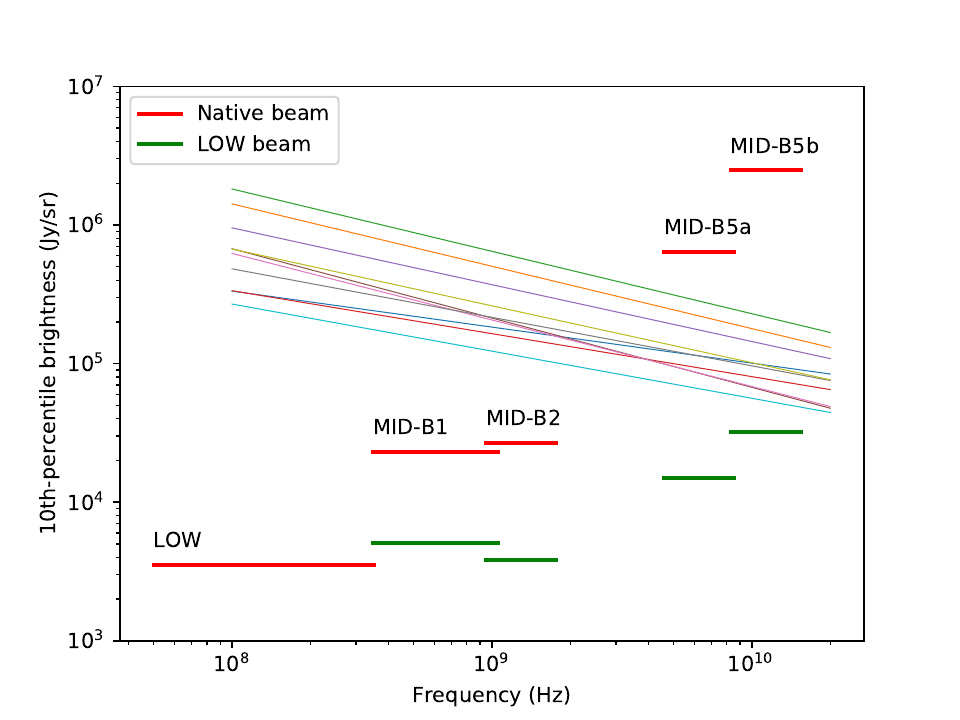}
    \caption{
    Simulated SEDs of the same 10 SNRs shown in Figure~\ref{fig:sens_mk} (narrow lines), compared with the SKA-LOW and SKA-MID sensitivity limits for a one-hour integration time, using native beam (red lines) or convolving to the SKA-LOW beam (green lines).}
    \label{fig:sens_low}
\end{figure}

First, we used the SKA sensitivity calculator to asses 
the capability of both LOW and MID 
for characterising Galactic SNRs SEDs (see Section~\ref{sec:sed}). As a representative example, we selected the 10 SNRs located closest to the Galactic centre reported 
by \citet{Loru2024}. 
To estimate the expected performance in terms of brightness, we calculated the 10th-percentile brightness level for each SNR in our sample from the SMGPS images used in \citet{Loru2024}, within the same masked regions adopted by those authors 
for the spectral index maps. 
The mean spectral index, $\alpha$, was then used to model the synchrotron emission over the  0.1--20~GHz range as a pure power law. In Figure \ref{fig:sens_mk}, we show the modelled SEDs of the 10 SNRs in our sample compared with the sensitivity of the SMGPS and the simulated SKA-MID sensitivity in Bands 1 and 2,
assuming a  one-hour 
integration time (roughly matching that of the SMGPS). For these simulations, we used all antennas available in the AA4 configuration (133 SKA dishes plus 64 
MeerKAT antennas) and set the robust parameter to 1.
When the SKA images are convolved to match the SMGPS beam, the achieved sensitivity improves by a factor of about 3 (dashed red lines). If we use the native beam for SKA, reaching a resolution of 
$1^{\prime\!\prime}.7$ and $1^{\prime\!\prime}.1$ for Bands 1 and 2, respectively (compared to $8^{\prime\!\prime}$ for the SMGPS), we can appreciate how SKA will be able to still detect low-brightness extended features even with a resolution improvement of almost one order of magnitude: this is an important step forward for SNRs, where fine structures (like arcs) co-exist with fainter diffuse emission.
Figure~\ref{fig:sens_low} compares the sensitivity limits for SKA-LOW and SKA-MID (for a one-hour integration, using robust$=0$ for SKA-LOW) with the SEDs of the 10 SNRs in our sample. It is worth noting that, in the SKA sensitivity calculator, natural weighting (which would represent a reasonable choice) is not yet fully supported, and that the total number of antennas available in Bands 5a and 5b is limited to 133.  

Since our main goal is to build spectral-index maps, we convolved the SKA-MID simulations to the SKA-LOW beam, obtaining 
the result shown as green lines in Figure~\ref{fig:sens_low}.
This approach yields an approximately uniform sensitivity (in Jy~sr$^{-1}$) across all SKA bands. We demonstrate that the SNRs in our sample, representative of a typical Galactic population in terms of brightness, can be fully imaged up to 15~GHz, provided that large-scale structure issues are properly accounted for (see Section~\ref{sec:sd}).

\begin{figure}
    \centering
    \includegraphics[height=8cm]{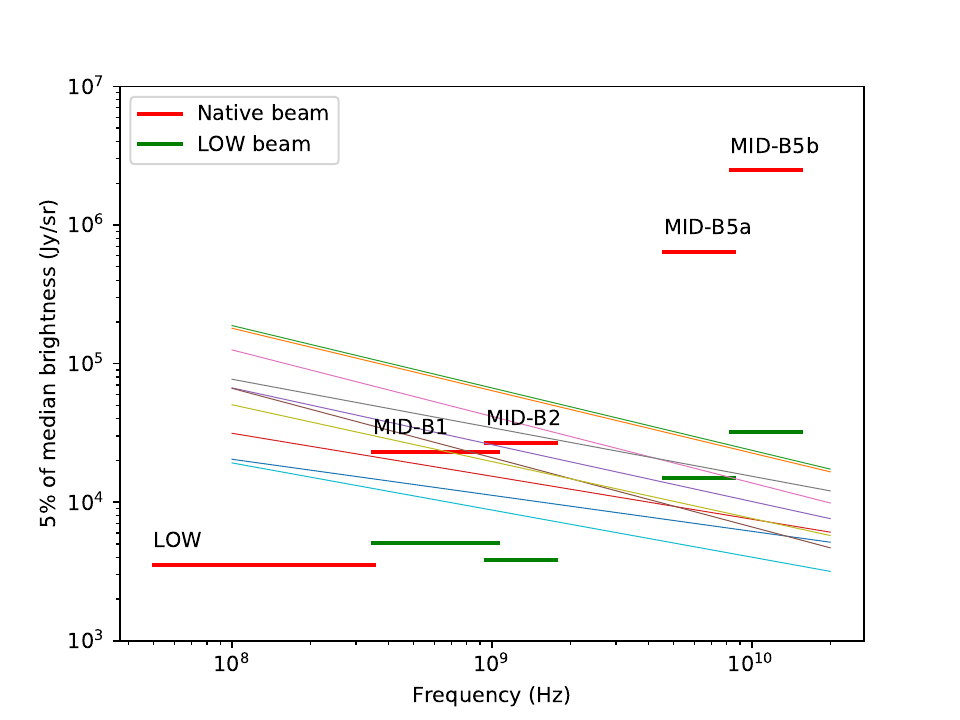}
    \caption{Same as Figure~\ref{fig:sens_low}, but considering 
    5\% of the median brightness value to simulate polarised emission.}
    \label{fig:sens_pol5}
\end{figure}

As discussed in Section~\ref{sec:pol}, polarisation maps are a fundamental tool for studying 
magnetic-field structures. With a typical polarisation fraction of 
5\%$-$25\%, the sensitivity limit found for Stokes-$I$ must be properly scaled to 
account for the corresponding reduction in polarised brightness. In Figure \ref{fig:sens_pol5},  we consider the same SNRs as in the previous figures, taking their median brightness (50th percentile) and multiplying this value 
by 0.05 to simulate polarised emission with a conservative 5\% polarisation fraction.

Using the native resolution for each band (with robust parameters discussed above) to 
achieve the highest possible resolution and to mitigate 
depolarisation introduced by 
beam averaging, the SKA will be able to detect the polarised emission from SNRs like 
those in our sample at frequencies up to SKA-MID Band 2 ($<1.76$~GHz). Polarised emission at the highest SKA frequencies (Band 5) 
will be difficult to observe even in well-known, brighter SNRs. 
It is worth noticing that this result is greatly affected by assumptions that may not hold in general: (i) the polarisation fraction may be higher than 5\%; (ii) the integration time may be increased for specific cases; (iii) depolarisation due to the Faraday rotation decreases at higher frequency (Band 5); and (iv) 
some SNRs are significantly brighter than those in our sample.


For spectral line observations, to detect the 1720-MHz OH masers, with a 10-minute integration, the SKA AA4 phase will achieve a sensitivity of $\sim$13~mJy for line emission at a velocity resolution of 0.15~km~s$^{-1}$ (for comparison, AA* sensitivity would be $\sim$20~mJy). This performance significantly exceeds the typical maser spot intensities (100--3000~mJy) detected in current VLBA or MERLIN observations \citep[e.g.,][]{Hoffman+2005a,Hoffman+2005b}. 


Table~\ref{tab:precision} summarises the astrometric precision achievable with SKA-VLBI for a $\sim$1~Jy source observed over five epochs, assuming
10-minute integrations per epoch and 
three distinct maser spots. 
The astrometric accuracy is estimated using two calibration techniques, namely phase referencing and MultiView. For phase referencing, we assume the reference quasar lies within two degrees of the target source (PR2$^{\circ}$). 
Following \citet{Rioja+2017}, who achieved approximately 0.1~mas positional precision in single-epoch $\sim$1.6~GHz OH maser observations, we adopt comparable precision estimates. 

\begin{table}[htbp]
  \footnotesize
  \centering
  \caption{Astrometric precision of SKA and SKA-VLBI, for different maximum baseline and beam size ($\theta_{\rm beam}$). A comparison between two calibration methods (PR2$^\circ$ and MultiView) is showed. For each method, the four columns report: the theoretical positional precision ($\sigma$); the thermal error of positional precision ($\sigma_{\rm pos,th}$); the final positional precision ($\sigma_{\rm pos}=\sqrt{\sigma^2+\sigma_{\rm pos,th}^2}$); the parallax precision, assuming that the target source exhibits three distinct maser spots, which were successfully observed across five separate epochs ($\sigma_\pi=\sigma_{\rm pos}/\sqrt{15}$).}
    \begin{tabular}{cccccccccccc}
    \hline \hline
    Telescope & Baseline & $\theta_{\rm beam}$ & \multicolumn{4}{c}{PR2$^{\circ}$}   &   & \multicolumn{4}{c}{MultiView} \\ \cline{4-7} \cline{9-12}
          &       &       & $\sigma$ & $\sigma_{\rm pos,th}$ & $\sigma_{\rm pos}$  & $\sigma_\pi$ & & $\sigma$ & $\sigma_{\rm pos,th}$ & $\sigma_{\rm pos}$ & $\sigma_\pi$ \\
          & (km) & (mas) & (mas) & (mas) & (mas) & (mas) & & (mas) & (mas) & (mas) & (mas) \\ \hline
    SKA AA* &       & $\sim$770  & 2.7   & $\sim$7.70 & $\sim$8.16 & $\sim$2.11 & & 0.1   & $\sim$7.70 & $\sim$7.70 & $\sim$1.98 \\
    SKA AA4 &       & $\sim$250  & 2.7   & $\sim$2.50 & $\sim$3.68 & $\sim$0.95 & & 0.1   & $\sim$2.50 & $\sim$2.50 & $\sim$0.65 \\
    SKA-VLBI & 1000  & $\sim$43   & 2.7   & $\sim$0.43 & $\sim$2.73 & $\sim$0.70 & & 0.1   & $\sim$0.43 & $\sim$0.44 & $\sim$0.11 \\
    SKA-VLBI & 2000  & $\sim$21   & 2.7   & $\sim$0.21 & $\sim$2.71 & $\sim$0.70 & & 0.1   & $\sim$0.21 & $\sim$0.23 & $\sim$0.06 \\
    SKA-VLBI & 3000  & $\sim$14   & 2.7   & $\sim$0.14 & $\sim$2.70 & $\sim$0.70 & & 0.1   & $\sim$0.14 & $\sim$0.17 & $\sim$0.04 \\
    \hline
    \end{tabular}%
  \label{tab:precision}%
\end{table}%


Using PR2$^{\circ}$, the positional precision improves to $\sim$2.7~mas, corresponding to a proper motion precision of $\sim$0.7~mas~yr$^{-1}$ over five epochs and a tangential-velocity precision of $\sim$3.3~km~s$^{-1}$ at a distance of $\sim$1~kpc, where 1~mas~yr$^{-1}$ $\approx$ 4.74~km~s$^{-1}$ at 1~kpc. 
Applying MultiView with 3,000~km baselines yields $\sim$0.17~mas positional precision and $\sim$0.04~mas~yr$^{-1}$ proper-motion accuracy, corresponding to a tangential-velocity precision of $\sim$0.18~km~s$^{-1}$ at $\sim$1~kpc.

If the AA4 configuration cannot be reached, SNR observations and the sensitivity limits described above will be moderately affected. For total-intensity observations, the decrease in sensitivity when using the AA* configuration poses the least concern for the study of known SNRs (which are usually very bright below $\sim\!1$~GHz), 
but it will significantly hinder the detection of new, faint, and extended SNR candidates. 
In polarisation studies, the loss of sensitivity is much more impactful, 
as significantly fewer SNRs would be detectable in Stokes $Q$ and $U$. 
Furthermore, the coarser resolution of the AA* configuration will reduce 
our ability to investigate small-scale filamentary structures and 
increase depolarisation effects due to beam averaging. 
The SKA AA* 
reaches $\sim$8~mas positional precision for 
1720~MHz~OH maser observations in SNRs, comparable to 
results obtained with the historical VLBA or MERLIN 
\citep{Hoffman+2005a,Hoffman+2005b}. 
Although several science cases could still be pursued with the AA* configuration, we emphasise the importance of completing the SKA Phase 1 according to its baseline design in order to fully    
exploit its scientific potential.

\subsection{Adding single-dish data to recover large structures}
\label{sec:sd}
Radio interferometers act like 
high-pass filters, being insensitive to spatial scales above a frequency-dependent threshold, which is mainly 
determined by the shortest spacing between the antennas. We 
refer to this threshold the largest angular scale (LAS), 
which can be approximated as 
\[
\theta_{\mathrm{LAS}}\sim\frac{\lambda}{b_{\mathrm{min}}},
\]
where $\lambda=c/\nu$ is the observation wavelength and $b_{\mathrm{min}}$ the interferometer's shortest baseline. As a result, quantitative and morphological information on sources more extended than, 
or comparable to, the LAS 
may be difficult to recover. 
A precise evaluation of the LAS is 
generally impossible, as it ultimately depends on the coverage of the inner part of the $uv$-plane, which, beyond the aforementioned dependence on frequency and projected shortest baselines, is also affected by the amount of flagged data and the weighting scheme. 
MeerKAT observations suggest that a reasonable value at $\sim\!1$~GHz is around half a degree \citep{Loru2024}. 
Scaled as $\nu^{-1}$, this value can also be assumed for SKA-MID. For SKA-LOW we expect a LAS greater than 1~degree, sufficient to properly image the vast majority of Galactic SNRs.

Providing information on spatial scales larger than the LAS is certainly crucial 
when studying a large sample of SNRs, whose dimensions span from a few arcminutes to a few degrees. The combination of interferometer data with 
single-dish telescope data is regarded as the standard procedure to overcome this limitation. 
The key idea is that a single-dish telescope 
of diameter $D$ probes the spatial scales larger than $\sim\lambda/D$, 
equivalent to providing visibilities in the $uv$-range from 0 to $\sim D/\lambda$. If 
a single-dish telescope with 
$D>b_{\mathrm{min}}$ is used, one can fully recover
the missing information from the inner part of the $uv$-plane. 
In practice, this simplistic solution faces two main challenges: (i) cross-calibrating the flux density scales between the interferometer and the single-dish telescope; (ii) deconvolving the single-dish data from the telescope beam. One of the most commonly used algorithms to perform the combination 
while addressing these requirements is ``feathering''. 
Schematically, the algorithm Fourier transforms the single-dish data, deconvolves them for the single-dish 
beam (by dividing the transformed data by the Fourier transform of the beam), and cross-calibrates the 
resulting single-dish ``visibilities'' with the 
interferometer visibilities. 
To ensure these operations can be properly performed, there should be 
sufficient overlap between the spatial scales probed by the two instruments. 
Although there is no strict limit for this overlap, given a shortest baseline for SKA-MID of approximately 30~m, the only suitable single-dish telescope in the southern hemisphere is currently Parkes. 

In this context, the Pegasus project is conducting observations with Parkes in the 0.7--4~GHz frequency range to provide zero-baseline information for ASKAP data. 
A full-sky map of the southern sky will be constructed. These data can also 
complement SKA-MID band-2 observations. 

Since the LAS scales 
with $\lambda$, the short-spacing problem 
becomes more severe at higher frequencies, potentially affecting the science proposed, especially in band 5.  Nevertheless, the Parkes telescope is planned to mount a new high-frequency receiver (UWH), expected in 2027, which 
will cover the frequency range of  SKA-MID bands 5a and 5b. 
Although an all-sky survey at these frequencies, similar to Pegasus, 
would be prohibitively time-consuming, targeted observations are certainly 
feasible for a reasonable number of SNRs 
requiring short-spacing information.

\section{Commensalities and synergies}
SNRs are among the few celestial objects that typically show detectable emission across the entire electromagnetic spectrum. In the following, we discuss possible commensalities with other SKA cases and synergies with other facilities.

\subsection{Commensalities with other SKA cases}

\subsubsection{SNRs in other galaxies}

Studying SNR populations in other galaxies 
offers significant advantages over Galactic studies by removing distance uncertainties and 
enabling the construction of complete population samples within individual galaxies. 
This allows statistical analysis of entire SNR populations 
and  facilitates the study of their collective impact on the host galaxy's properties and evolution. However, these benefits come with challenges: the greater distances of extragalactic systems reduce both sensitivity and spatial resolution, and introduce selection biases favouring only the brightest remnants.

The enhanced observational capabilities of the SKA will help mitigate these limitations. This potential is already evident from the achievements of its precursors in mapping the SNR populations of the Magellanic Clouds. 
The Small Magellanic Cloud (SMC) has 
been the target of multiple surveys across different frequencies \citep{Filipovic2005,Payne2007,Filipovic2008,Maggi2019,2019MNRAS.490.1202J,Leahy2022}. The most recent census, based on new MeerKAT observations \citep{Cotton2024}, catalogued the entire known population of 21 SNRs, discovered 3 new remnants, and identified 10 additional candidates. Similarly, the LMC has been extensively surveyed at several frequencies \citep{Maggi2016,Bozzetto2017,Yew2021,Bozzetto2023,Zangrandi2024}, with the most recent inventory listing 77 confirmed SNRs and 47 candidates \citep{Zangrandi2024}.
The SKA precursors have been instrumental in these studies. MeerKAT has provided the most complete SMC sample to date, while ASKAP has produced the best current radio data for the LMC \citep{Pennock2021}, which led to the identification of 1 new SNR and 13 new candidates \citep{Bozzetto2023}. In addition, MeerKAT observations have enabled detailed analyses of individual LMC remnants, including the discovery of the first circumgalactic SNR,  J0624$-$6948 \citep{Filipovic2022,Sasaki2025}. Particularly in th case of the LMC, several of the candidates still await confirmation through detection at multiple frequencies--a task for which the SKA, with its greatly improved survey sensitivity, is uniquely well suited. 

While the Magellanic Clouds remain the prime targets due to their proximity, surveys have also been conducted in more distant galaxies. Examples include the nearby M31 and M33 in the Local Group \citep{Galvin2014_M31,Gordon1999}, as well as other galaxies such as M82, NGC~6946, M51, and members of the Sculptor Group \citep{Huang1994,Fenech2008,Lacey1997,Maddox2007,Pannuti2000,Pannuti2002,Galvin2012,OBrien2013,Galvin2014}. However, sensitivity and angular resolution constraints currently limit the detection of SNRs at larger distances, and the number of radio-detected remnants decreases rapidly beyond the Local Group \citep{Urosevic2005}. The SKA will overcome many of these challenges. Its unprecedented sensitivity and resolution will enable detailed imaging and characterisation of SNRs in more distant galaxies, improving detection statistics and reducing selection biases in existing samples. This capability will open the way for comprehensive population studies across a broader range of galactic environments, providing new insight into the interplay  between SNR evolution, star formation, and galactic feedback processes. 

\subsubsection{Galactic plane survey with SKA-MID Band 5}
Among the science cases proposed for SKA-MID, a Galactic plane survey has been designed using the Band-5b receivers (10--15~GHz; \citealt{Traficante01.2026.SKA}). Although 
this survey is primarily intended 
to detect 
thermal emission from star-forming regions, 
it will also offer significant opportunities for SNR studies. While SNRs are generally easier to detect at lower frequencies, a Band-5 survey represents a valuable complement to existing datasets and can be exploited for several purposes. 
First, as noticed in Section~\ref{sec:calc}, the sensitivity of SKA-MID in the AA4 configuration 
will be sufficient to detect a large fraction of known SNRs, 
for which data at these frequencies are often missing. 
With an expected sensitivity of approximately 20~$\mu$Jy and an angular resolution of about $0^{\prime\prime}.8$ at 15~GHz, 
the survey will allow the measurement of 
spatially resolved spectral indices with a resolution of roughly 500~AU at a distance of 20~kpc.
This capability will 
make it possible to characterise 
both 
global and local variations in the spectral index across hundreds of Galactic SNRs, providing crucial information on the underlying electron energy distribution and synchrotron ageing effects. 
Second, high-frequency observations also bring clear advantages for polarisation studies. Depolarisation due to Faraday rotation and beam averaging decreases rapidly with frequency, enabling more accurate mapping of the intrinsic magnetic field structure within the remnants. Combined with SKA's high dynamic range and sensitivity, Band-5 data will thus be instrumental in tracing the geometry and strength of magnetic fields, which are key to understanding shock compression and particle acceleration processes.
Finally, a blind survey at these frequencies could 
lead to the discovery of new 
SNRs.
While most remnants are traditionally identified at lower frequencies, recent results from SKA precursors (e.g., \citealt{Anderson2025}) demonstrate that previously undetected or confused remnants can still emerge at higher frequencies.

\subsection{Synergies with other facilities}

\subsubsection{Sardinia Radio Telescope}
Synergy with new-generation single-dish radio telescopes, such as the Sardinia Radio Telescope (SRT), offers 
valuable opportunities to explore the high-frequency ($>20$~GHz) spectral behaviour of SNRs, 
effectively complementing the frequency coverage of the SKA. 
Observations in this regime enable the identification of 
potential spectral cut-off associated with the maximum energy of accelerated CR electrons, thereby 
placing firmer constrains 
on the distinction between purely leptonic and hybrid lepto-hadronic models. $K$-band observations performed with the SRT on extended SNRs 
such as I~C443 and W44 \citep{Loru2021} have already demonstrated the instrument's excellent trade-off between sensitivity and resolution. 
Major progress in this field are expected with the implementation of the new tri-band receiver \citep{Bolli_2023}, which will enable simultaneous observations 
in three frequency bands, $18$--$26$~GHz ($K$), $34$--$50$~GHz ($Q$), and $80$--$116$~GHz ($W$; \citealt{Navarrini_2022}). These forthcoming  observations will allow for more detailed and higher-frequency studies of the non-thermal emission from 
CR electrons, significantly 
enhancing our ability to investigate spectral trends and 
potential cut-offs associated with the maximum energies of accelerated particles.

\subsubsection{Imaging X-ray Polarimetry Explorer}
\label{sec:ixpe}
Since the launch of 
IXPE in 2021, 
polarised synchrotron emission has also been 
detected in X-rays. The orientations of 
the polarisation vectors in SNRs 
at  X-ray energies generally correspond to those observed in the radio band \citep{2024Galax..12...59S}. 
The measured polarisation fractions are typically higher in X-rays, as expected. Indeed, 
the maximum polarisation fraction, given by  $(s+1)/(s+7/3)$ increases with the spectral index $s$, 
which is larger for electrons emitting in X-rays 
due to the high-energy cut-off in 
their energy distribution. 

IXPE and the SKA 
probe different portions of the electron energy spectrum, and 
their combined observations provide a powerful diagnostic of 
magnetic-field topology across scales--from compact regions 
traced in X-rays to extended lobes observed in 
radio. 
This synergy will also improve our understanding of particle acceleration to different energies and of the behaviour of magnetic turbulence over a broad range of sales.

\subsubsection{Cherenkov Telescope Array Observatory}
\label{sec:cta}
New $\gamma$-ray observations with the next Imaging Atmospheric Cherenkov Telescopes (IACTs), such as 
the CTAO, 
will be crucial 
for obtaining high-resolution images of extended SNRs, allowing for in-depth studies of 
morphology and spectra 
at sub-arcminute resolution across TeV energies. At these energies, the hadronic contribution, if present, is expected to 
become evident in the SNR SED, 
providing a crucial observational window to disentangle 
leptonic and hadronic emission components. Moreover, 
above several tens of GeV, a dominant inverse Compton 
contribution from 
electrons is expected in the $\gamma$-ray spectrum \citep{Loru2021}. Observations with the CTAO 
will thus provide additional constraints which, when combined with radio data, will shed new light on the electron contribution to the high-energy spectral features of SNRs

\subsubsection{Cubic Kilometre Neutrino Telescope}
\label{sec:km3net}
The Cubic Kilometre Neutrino Telescope (KM3NeT)\footnote{\url{https://www.km3net.org/}} is a next-generation neutrino 
observatory. The SKA could serve as a 
complementary instrument for identifying potential counterparts to  neutrino detections within the neutrino error regions,  
as well as for monitoring multi-messenger alerts.

Since the particles responsible for radio emission (electrons) differ from those producing neutrinos (protons), a joint analysis of KM3NeT and SKA data could help disentangle the hadronic and leptonic components 
of $\gamma$-ray emission from the same source. 
Moreover, while $\gamma$-rays  can be absorbed by the surrounding medium, radio waves propagate freely, making radio observations a crucial tool for characterising the environments of potential neutrino sources. In particular, 
such observations can trace the high-density structures that are essential for the production of hadronic $\gamma$-rays and neutrinos.
These combined studies could be especially relevant for SNRs during the immediate post-explosion phase.










\section*{Acknowledgements}
DL acknowledges the support by the National SKA Program of China (grant No.~2022SKA0120103) and the NSFC Grants No.~12503071.

\bibliographystyle{abbrvnat-maxbibnames4}
\bibliography{chapter} 

\end{document}